\documentclass[prb,twocolumn,showpacs,superscriptaddress,amsmath,amssymb]{revtex4} 
\usepackage{graphicx} 
\usepackage{amsmath} 
\usepackage{color}

\long\def\comment#1{}
\begin{document}

\title{Exactly solvable Kitaev model in three dimensions}

\author{Saptarshi Mandal}
\affiliation{ The Institute of Mathematical Sciences, C.I.T
  Campus, Chennai 600113, India}
\email{saptarshi@imsc.res.in}
\author{Naveen Surendran} 
\affiliation{The Abdus Salam International Centre for
  Theoretical Physics, Strada Costiera 11, 34014 Trieste, Italy}  
\email{nsurendr@ictp.it}
\date{\today}

\begin{abstract}
  We introduce a spin-$\frac{1}{2}$ model in three dimensions which is
  a generalization of the well-known Kitaev model on a honeycomb
  lattice. Following Kitaev, we solve the model exactly by mapping it
  to a theory of non-interacting fermions in the background of a
  static $\mathbb{Z}_2$ gauge field. The phase diagram consists of a
  gapped phase and a gapless one, similar to the two-dimensional case.
  Interestingly, unlike in the two-dimensional model, in the gapless
  phase the gap vanishes on a contour in the ${\bf k}$ space.
  Furthermore, we show that the flux excitations of the gauge field,
  due to some local constraints, form loop like structures; such loops
  exist on a lattice formed by the plaquettes in the original
  lattice and is topologically equivalent to the pyrochlore lattice.
  Finally, we derive a low-energy effective Hamiltonian that can be
  used to study the properties of the excitations in the gapped phase.

\end{abstract}
\maketitle

\section{Introduction}

The study of topological phases has been actively pursued in
condensed-matter systems for some years. This has resulted in the
emergence of a new paradigm in the theory of quantum phase
transitions: certain phase transitions cannot be described in terms of
local order parameters associated with spontaneously broken
symmetries; instead, the phases in such transitions are characterized
by topological order.\cite{Wen1}  The most famous example of
topological order in a quantum system is in the phenomenon of
fractional quantum Hall effect.\cite{Laughlin,WenNiu,MooreRead} Other
examples---experimental and theoretical---include quantum spin
liquids,\cite{BaskaranZou,WenWilczek,ReadSachdev,Wen2} quantum dimer
models,\cite{RokhsarKivelson,ReadChakraborty,MoessnerSondhi} quantum
loop models,\cite{Freedman} etc.

Recently, it has been proposed that topological phases can be used to
do quantum computation.\cite{Shor,Kitaev1} The main obstacle in the
realization of quantum memory---the basic ingredient of a quantum
computer---is decoherence: it is difficult to prepare states that are
robust to external noise.  Kitaev\cite{Kitaev2} suggested that
topologically ordered states can be used to overcome this problem.  He
illustrated these ideas in a spin-$\frac{1}{2}$ model on a hexagonal
lattice which, quite remarkably, can be solved exactly. The ground
state of the Kitaev model has two phases: in one phase the elementary
excitations have a gap in the spectrum and are Abelian anyons; the
second phase is gapless in the absence of an external magnetic field,
but develops a gap when the field is switched on, and then the
excitations are non-Abelian anyons.  In topological quantum
computation, braiding of non-Abelian anyons is essential for the
realization of universal quantum gates.\cite{NayakSimon} Regarding the
feasibility of a physical realization of the Kitaev model, there has
been a proposal to realize it on an optical
lattice.\cite{DuanDemler,MicheliBrennen}

Apart from its potential application in quantum computation, the
Kitaev model is an interesting many-body system by itself. First,
exact solutions are rare in dimensions higher than one, and second,
the model provides a relatively simple platform---the Hamiltonian
involves only two-body interactions---to study concepts such as
topological order and fractional excitations. Thus, not surprisingly,
the various many-body aspects of the model have been thoroughly
investigated.
\cite{BaskaranMandal,ChenNussinov,ChenHu,SchmidtDusuel,DusuelSchmidt,SenguptaSen}
There have also been some generalizations to other two-dimensional (2D)
lattices.\cite{YangZhou,YaoKivelson} It is then worthwhile to find
generalizations of the model in higher dimensions. In this paper, we
introduce and study a three-dimensional (3D) version of the Kitaev model.
Three-dimensional models exhibiting topological order have been
studied previously.\cite{LevinWen,HammaZanardi,BombinDelgado1,BombinDelgado2}

An exact solution of the Kitaev model is possible due to the existence
of a macroscopic number of locally conserved quantities: this
facilitates the mapping of the model to a quadratic Hamiltonian of
Majorana fermions hopping in the background of a static $\mathbb{Z}_2$
gauge field. The three-dimensional model we construct also has the
above feature, which renders it exactly solvable. It has a gapped
phase and a gapless one, just as in 2D, and although the phase
boundaries are identical to the latter, the nature of certain
excitations is quite different. In the three-dimensional model, the
excitations of the gauge field are localized on ``loops''. This gives
rise to the possibility that, as yet unverified in our model, such
excitations can obey nontrivial statistics, since double exchange in
loops---unlike points---is topologically nontrivial in three
dimensions (for example, see Ref. \onlinecite{BombinDelgado1}).

The paper is organized as follows: in Sec. \ref{sec-hamiltonian} we
define the spin Hamiltonian and then rewrite it using a Majorana
fermion representation of spin $\frac{1}{2}$. In Sec.
  \ref{sec-spectrum} the low-energy spectrum for fermionic
  excitations is derived while Sec. \ref{sec-excitations}
  discusses the excitations of the gauge field. We then derive the
  low-energy effective Hamiltonian in the large $J_z$ limit in Sec.
  \ref{sec-heff} and end with a discussion of our results in the
  last section.

\section{\label{sec-hamiltonian} Hamiltonian}

\begin{figure}
\includegraphics[width=.4\textwidth]{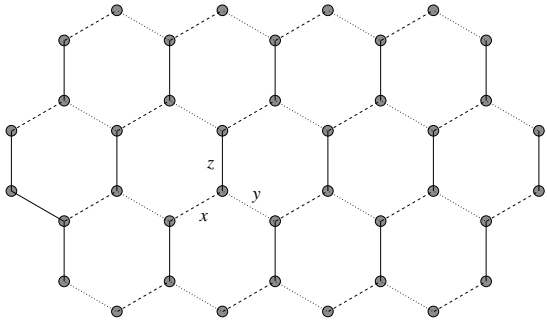}
\caption{\label{hexlat} The honeycomb lattice: the three 
  types of links are labeled $x$, $y$, and $z$.}
\end{figure} 
Before presenting the three dimensional model, let us briefly describe
the Kitaev model in 2D. It is a system of spin-$\frac{1}{2}$
degrees of freedom located at the vertices of a honeycomb lattice.
In the honeycomb lattice, there are three types of links which,
distinguished by their orientations, are labeled $x$, $y$ and $z$ (see
Fig. \ref{hexlat}).  The Kitaev Hamiltonian is
\begin{equation}
\label{kitaevham}
H = -J_x \sum_{<i,j>_x} \sigma_i^x \sigma_j^x - J_y \sum_{<i,j>_y} \sigma_i^y
\sigma_j^y -J_z \sum_{<i,j>_z} \sigma_i^z \sigma_j^z,
\end{equation}
where $\sigma^a$'s are the Pauli matrices, and  $<i,j>_a$ indicates that $i$
and $j$ belong to a link of $a$-type.

For our purposes, the key points to note about the Hamiltonian in Eq.
(\ref{kitaevham}) are that for any spin, only one of the components
couples to a particular neighboring spin, and the component with
non-zero coupling strength is different for each of the three
neighbors. As we will see later, this leads to the existence of a set
of mutually commuting conserved plaquette operators, which in turn
makes the problem exactly solvable.

We note two features of the honeycomb lattice that are pertinent to
the construction of the 3D lattice: (1) The coordination number of the
lattice is 3.  (2) The three types of links $x,y$ and $z$ are
distributed in such a way that two links of the same type do not touch
each other.

\subsection{\label{sec-3dlattice}Lattice}

To facilitate visualization, we will first describe how to obtain the
3D lattice starting from the familiar cubic lattice. Let $i,j,k \in
\mathbb{Z}$ be the $x,y$ and $z$ coordinates of the latter. The new
lattice is obtained by removing those sites that satisfy one of the
following conditions: (1) $k = 0 \bmod 4$ and $i = 0 \bmod 2$, (2) $k
= 1 \bmod 4$ and $j = 0 \bmod 2$, (3) $k = 2 \bmod 4$ and $i = 1 \bmod
2$, and (4) $k = 3 \bmod 4$ and $j = 1 \bmod 2$.

\begin{center}
\begin{figure}[ht]
\includegraphics[width=.42\textwidth]{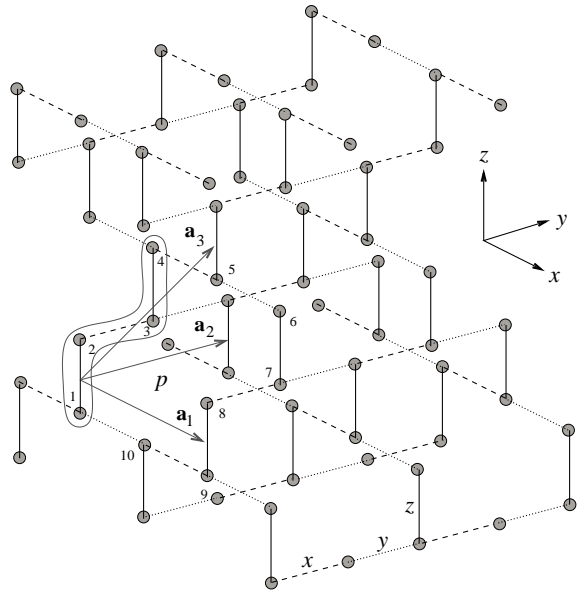}
\caption{ \label{fig-3dlattice} The 3D lattice: the four sites inside
  the loop (marked 1-4) constitute a unit cell; ${\bf a}_1, {\bf
    a}_2,$ and ${\bf a}_3$ are the basis vectors. Plaquette
  $p$ consists of sites marked 1-10.}
\end{figure}
\end{center}

This amounts to depleting the cubic lattice by half, and the resultant
lattice has coordination number 3 (see Fig. \ref{fig-3dlattice}). We
note that: (i) The $x$-$y$ planes alternately consist of disconnected
rows or disconnected columns. (ii) As one goes along a particular row
(column), at each site there is a link whose direction alternates
between positive and negative $z$ axes. That is, there is a
criss-crossing structure between adjacent planes which ensures that
the lattice is truly three dimensional---despite a coordination number
of 3---and not a set of mutually disconnected two-dimensional
surfaces.

To parametrize the lattice sites, we first note that the unit cell
contains four sites. The position vector of a unit cell is given by
\begin{eqnarray}
{\bf R} &=& m {\bf a}_1 + n {\bf a}_2 + p {\bf a}_3,~~m,n,p \in
\mathbb{Z}, \label{unitcell-pos} \\
{\bf a}_1 &=& 2 \hat {\bf x},~~{\bf a}_2 = 2 \hat {\bf y},~~
{\bf a}_3 =\hat {\bf x} + \hat {\bf y} +  2 \hat {\bf z},
\label{basis}
\end{eqnarray}
where $\hat {\bf x}, \hat {\bf y}$, and $\hat {\bf z}$ are unit vectors
along $x,y$ and $z$ directions, respectively.  The four sites within a
unit cell are at
\begin{eqnarray}
{\bf r}_1 = {\bf R} -\frac{\hat {\bf y}}{2}-
\hat {\bf z},&& {\bf r}_2 = {\bf R} -\frac{\hat {\bf y}}{2},\nonumber
\\
{\bf r}_3 = {\bf R} +\frac{\hat {\bf y}}{2},&& 
 {\bf r}_4 = {\bf R}+\frac{\hat {\bf y}}{2}+
\hat {\bf z}.
\label{sites-in-ucell}
\end{eqnarray}
To define a Kitaev-type Hamiltonian, we need one more ingredient, viz,
the labeling of links. To this end, we alternately assign $x$ and $y$
labels to the links in each of the rows and columns that lie on the
$x$-$y$ plane; the remaining links, the ones along the $z$ axis, are
labeled $z$.  [The ambiguity in the assignment of $x$ and $y$ labels
within each row (column) is resolved by demanding periodicity.] This
way of labeling ensures that the three links emanating from each site
have different labels. Now the definition of the Hamiltonian in Eq.
(\ref{kitaevham}) can be applied to the three-dimensional lattice we
have constructed. Explicitly,
\begin{widetext}
\begin{eqnarray}
H &=& \sum_{{\bf R}} \Big[-J_x \sigma^x_{1}({\bf R}) 
\sigma^x_{4}({\bf R} - {\bf a}_3)  
-J_y \sigma^y_{1}({\bf R}) 
\sigma^y_{4}({\bf R} + {\bf a}_1-{\bf a}_3) 
-J_x \sigma^x_{2}({\bf R}) 
\sigma^x_{3}({\bf R}) \nonumber \\
&&~~\qquad - J_y \sigma^y_{3}({\bf R}) 
\sigma^y_{2}({\bf R} + {\bf a}_2) 
-J_z \sigma^z_{1}({\bf R}) \sigma^z_{2}({\bf R})
-J_z \sigma^z_{3}({\bf R}) \sigma^z_{4}({\bf R}) \Big].
\label{3dham}
\end{eqnarray}
\end{widetext}
For simplicity of notation, we will continue to use the formal
expression for $H$ in Eq. (\ref{kitaevham}), where sites are referred to by a
single lower-case index, and revert to the explicit form in Eq. (\ref{3dham})
only when the calculation demands it. 

\subsection{\label{sec-conserved} Conserved quantities}
Let $l=(j_1, j_2, \dots, j_n)$ be a sequence of lattice sites such
that $j_m$ and $j_{m+1}$ are neighbors for $m=1,2, \dots n$ with
$j_{n+1} \equiv j_1$. Then $l$ represents a loop. Equivalently, the links
formed among $j_m$ also uniquely determine the loop. Let 
\begin{equation}
W_l = \prod_{j_m\in l} \sigma^{a_{j_m}}_j,
\label{loop}
\end{equation}
where $a_{j_m}$ is
the label of the link that connects $j_m$ to its neighbor which is
not in $l$.  It is easy to see that 
\begin{equation}
\label{conservedqts}
[W_l, H] = 0,~  [W_l, W_{l'}] = 0,~ 
\forall l,l'.
\end{equation}

However, not all such operators are independent. To see this, we
define a geometrical operation of combining two loops as follows: the
combination of loops $l_1$ and $l_2$ is the loop $l_{12}$ formed by all the
links in $l_1$ and $l_2$ except those that are common to both. Then,
it is easy to see that $W_{l_1} W_{l_2} = \pm W_{l_{12}}$. This means that
for a lattice with open boundary conditions, any loop operator can be
written as a product of those defined on the plaquettes---the elementary
loops which cannot be obtained by combining smaller
loops. For example, in Fig. \ref{fig-3dlattice}, $W_p$ corresponding to the
plaquette $p$, consisting of the sites marked $1, 2, \dots, 10$, is
\begin{equation}
W_p = \sigma_1^x \sigma_2^y \sigma_3^y \sigma_4^y \sigma_5^z
\sigma_6^x \sigma_7^y \sigma_8^y \sigma_9^y \sigma_{10}^z.
\label{plaquetteop}
\end{equation}
There are further constraints among $W_p$ corresponding to different
plaquettes, and we will discuss them in Sec.
\ref{sec-excitations}, where the excitations are studied.

\subsection{Hamiltonian in Majorana fermion representation}

To solve the Hamiltonian, following Kitaev,\cite{Kitaev2} we use a
representation of Pauli matrices in terms of Majorana fermions. At
each site $j$, we introduce four Majorana operators $b_j^x$, $b_j^y$,
$b_j^z$ and $c_j$, which satisfy the following relations.
\begin{eqnarray}
{b^\alpha_j}^\dagger = b^\alpha_j,&& c_j^\dagger = c_j, \nonumber \\
\{b^\alpha_j,b^\beta_k\} = 2 \delta_{jk} \delta_{\alpha \beta}, &&
\{ c_j, c_k \} = 2\delta_{jk}, \nonumber \\
\{b^\alpha_j,c_l\} &=&0.
\label{majorana}
\end{eqnarray}
 The Majorana fermions act on a Fock space which is four
dimensional, whereas the spin Hilbert space has two dimensions. We define
the physical space as consisting of those states which satisfy the
constraint
\begin{equation}
  D_j|\xi \rangle = |\xi \rangle,~~\forall j~~\mbox{where } D_j =
  b_j^x b_j^y b_j^z c_j. 
\end{equation}
$D_j^2=1$; thus, $(1, D_j)$ form the elements of a $\mathbb{Z}_2$
gauge group. The spin operators are defined as
\begin{equation}
\sigma_ j^\alpha = i b_j^\alpha c_j .
\label{spin-majorana}
\end{equation}
When restricted to the physical space, the above operators satisfy the
standard spin-$\frac{1}{2}$ algebra. The projector to the physical
space is given by
\begin{equation}
P_j = \frac{1+D_j}{2}.
\label{projector}
\end{equation}
All the states related by a gauge transformation project on to the
same physical state.

Starting from a generic spin model, $H\{\sigma_j^\alpha\}$, we can
obtain a fermionic Hamiltonian, $\widetilde H\{b_j^\alpha, c_j\}$, by
using Eq. (\ref{spin-majorana}). Since $[\widetilde H,P_j] =
0~\forall j$, the eigenstates of $H$ can be obtained from those of
$\widetilde H$ by projecting the latter to the physical space.

Substituting Eq. (\ref{spin-majorana}) in Eq. (\ref{kitaevham}), we
obtain 
\begin{eqnarray}
\widetilde H&=& \frac{i}{2} \sum_{j,k} \hat A_{jk} c_j c_k, \\ 
\hat A_{jk}&=& \left\{ 
\begin{array}{cl}
J_{\alpha_{jk}}  \hat u_{jk} & \mbox{if $j$ and $k$ are linked,} \\
0 & \mbox{ otherwise,}
\end{array}
\right. \label{Ajk}\\
\hat u_{jk} &=& i b_j^{\alpha_{jk}} b_k^{\alpha_{jk}},
\label{link-conserved}
\end{eqnarray}
where $\alpha_{jk}$ is the type of the link between $j$ and $k$.  We
note that $ \hat u_{jk} = - \hat u_{kj} $, and in the sum the links
are treated as directed and therefore counted twice. We use a hat to
emphasize that $\hat u_{jk}$ is an operator; $u_{jk}$ is the
corresponding eigenvalue and takes values $\pm1$.  Furthermore,
\begin{equation}
[\widetilde H, \hat u_{jk}] = 0,~ [\hat u_{jk}, \hat u_{lm}] =0~
\forall~ <j,k>,<l,m>.
\label{ujk}
\end{equation}
Therefore, the Hilbert space breaks up into various sectors, each
corresponding to a particular set $\{u_{jk}\}$; the matrix elements of
$\widetilde H$ between states belonging to different sectors are zero.
The Hamiltonian in a given sector is obtained by replacing the
operators $\hat u_{jk}$ with their corresponding eigenvalues, $u_{jk}$: 
\begin{equation}
\widetilde H_u = \frac{i}{2} \sum_{j,k} A_{jk} c_j c_k,
\label{hamu}
\end{equation}
where $A_{jk}$ is obtained from Eq. (\ref{Ajk}) by substituting $\hat
u_{jk}$ with $u_{jk}$. The gauge-invariant and hence physical conserved
quantities are 
\begin{equation}
\widetilde W_p = \prod_{m=1}^{10} \hat u_{j_mj_{m+1}},
\label{physical-conserved}
\end{equation}
where $p=(j_1,j_2, \dots,j_{10})$, as before, is a plaquette.
$\widetilde W_l$ is related to $W_l$ in Eq. (\ref{loop}) as follows:
\begin{equation}
W_l = \mathcal{P}_l \widetilde W_l \mathcal{P}_l,~~\mathcal{P}_l =
\prod_{j \in l} P_j.
\end{equation}
From now on we will simplify the notation, following
Kitaev,\cite{Kitaev2} by not making distinction between operators in
the physical and extended Hilbert spaces, i.e., we will drop tilde
from all the operators acting in the latter.

\section{\label{sec-spectrum} Ground state and spectrum} 

Next question to be addressed is: Which sector of $\{u_{ij}\}$ does
the ground state belong to? The problem of free fermions hopping on a
$d$-dimensional hypercubic lattice with hopping amplitude $|t_{ij}|
e^{i\theta_{ij}}$ between nearest-neighbor sites $i$ and $j$ was
studied by Lieb.\cite{Lieb} Moreover it was shown that, if $|t_{ij}|$ is
reflection symmetric about certain planes that does not contain any
sites, then the ground-state energy is the lowest when the flux of the
phase along the plaquettes, $\Phi \equiv \sum_{<ij>} \theta{ij}$,
equals $\pi$ if the length of the loop is 0 mod 4, and zero if the
length is 2 mod 4. Unfortunately, Lieb's result cannot be directly
applied to our case because our lattice does not have the required
reflection symmetry. 

We numerically studied the ground-state energy of the Hamiltonian in
Eq. (\ref{hamu}) for lattices containing up to 864 sites using
periodic boundary conditions. The ground-state energy in the sector
with uniform flux ($W_p=+1, \forall p$) was compared with that of
other selectively chosen flux configurations. As will be discussed in
Sec. \ref{sec-excitations}, due to some local constraints, the
plaquettes that are excited themselves form loops in an embedded
lattice. The smallest such loop consists of six plaquettes. We
considered excitations of loops of varying length and found the
energy increasing with increasing length. We also looked
at excitations of multiple loops. In every case considered, we found
that the energy of the lowest energy state is greater than that of
the $W_p=+1$ uniform flux state.  In Sec. \ref{sec-heff}, we will get
further confirmation of this result, at least in the large $J_z$
limit, through a perturbative analysis. Therefore it is reasonable to
assume that the ground state belongs to this sector.

\begin{center}
\begin{figure}
\includegraphics[width=.4\textwidth]{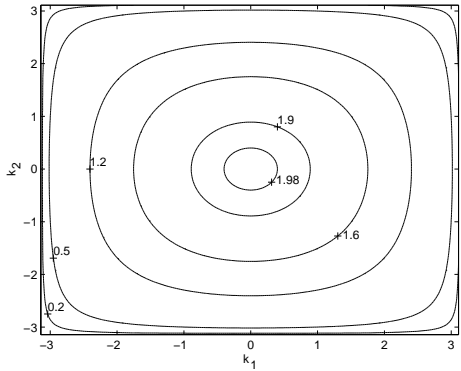}
\caption{ In the ${\bf k}$ space, the contour on which the gap
  vanishes is projected on $k_3=0$ plane. $J_x=J_y=1$, and $J_z$ is
  varied between zero and two. Corresponding values of $J_z$ are shown next
  to the contours. }
\label{fermisurface}
\end{figure}
\end{center}

$u_{jk}=1$ for all links is the obvious choice among the
configurations of link variables that gives $W_p=+1, \forall p$. Of
course, any configuration related to this one by a gauge
transformation will also satisfy the above condition on $W_p$'s. With
this choice, the Hamiltonian in its explicit form becomes
\begin{widetext}
\begin{eqnarray}
H &=& i\sum_{{\bf R}} \Big[J_x c_{1}({\bf R}) 
c_{4}({\bf R} - {\bf a}_3)  
+J_y c_{1}({\bf R}) 
c_{4}({\bf R} + {\bf a}_1-{\bf a}_3) 
+J_x c_{2}({\bf R}) c_{3}({\bf R}) \nonumber \\
&& \qquad ~~ + J_y c_{3}({\bf R}) c_{2}({\bf R} + {\bf a}_2) 
+J_z c_{1}({\bf R}) c_{2}({\bf R})
+J_z c_{3}({\bf R}) c_{4}({\bf R}) \Big]
\label{gssectorham}
\end{eqnarray}
\end{widetext}
where ${\bf r}$ and ${\bf a}_i$ are given in Eqs.~(\ref{unitcell-pos})
and (\ref{basis}). To diagonalize the Hamiltonian, we next do a Fourier
transform.
\begin{equation}
c_{\mu}({\bf r}) = 
 \int_{-\pi}^{\pi} \frac{dk_1}{2\pi} 
\int_{-\pi}^{\pi} \frac{dk_2}{2\pi}
\int_{-\pi}^{\pi}  \frac{dk_ 3}{2\pi}
e^{-i {\bf k}\cdot {\bf r}}
  c_{\mu\alpha}({\bf k}),
\end{equation}
with $\mu =1,2,3,4$ and where
\begin{eqnarray*}
 {\bf k} &=& k_1 {\bf b}_1 + k_2 {\bf b}_2 + k_3 {\bf b}_3,
\end{eqnarray*}
and
\begin{eqnarray*}
&& {\bf b}_1 = \frac{(2 \hat {\bf x} - \hat {\bf z})}{4} ,~~
 {\bf b}_2 = \frac{(2 \hat {\bf y} - \hat {\bf z})}{4} ,~~
 {\bf b}_3 = \frac{\hat {\bf z}}{2} . 
\end{eqnarray*}
Using the property, $c_{\mu}(-{\bf k}) = 
c_{\mu}^\dagger({\bf k})$, the Hamiltonian becomes,
\begin{widetext}
\begin{eqnarray}
H&=& \int_{-\pi}^{\pi} \frac{dk_1}{2\pi} 
\int_{-\pi}^{\pi} \frac{dk_2}{2\pi}
\int_{-\pi}^{\pi}  \frac{dk_ 3}{2\pi} \Bigg[ 
\frac{i}{2} \bigg\{ e^{ik_3} \delta_{k_1} 
c_{1}^\dagger({\bf k}) c_{4}({\bf k})
+ \delta_{k_2} c_{3}^\dagger({\bf k}) c_{2}({\bf k})
+ J_z c_{1}^\dagger({\bf k}) c_{2}({\bf k}) \bigg\} + h.c. \Bigg],
\label{kham} 
\end{eqnarray}
where $\delta_{k_i} = J_x + e^{-ik_i} J_y$, for $i=1,2$.
Furthermore, we define \mbox{$\Delta_{\bf k} = \big(|\delta_{k_1}|^2 + 
|\delta_{k_1}|^2 + 2 J_z^2
\big)$}, and $\phi_{\bf k}$ such that \mbox{$e^{-ik_3} \delta_{k_1} 
\delta_{k_2} \equiv |\delta_{k_1}|  |\delta_{k_2}| e^{i\phi_{\bf k}}$}. 
The above Hamiltonian is easily diagonalized and we obtain the
spectrum to be
\begin{equation}
  E({\bf k}) = \pm \frac{1}{2\sqrt{2}} \Bigg[ \Delta_{\bf k} \pm \bigg[ \Delta_{\bf k}^2 -
  \Big\{\big(J_z^2 - |\delta_{k_1}| |\delta_{k_2}|\big)^2 
  + 2 J_z^2 (1-\cos \phi_{\bf k}) |\delta_{k_1}| |\delta_{k_2}| 
  \Big\}\bigg]^{\frac{1}{2}} 
  \Bigg]^{\frac{1}{2}}. \label{spectrum}
\end{equation}
\end{widetext}
In the ground state, all the negative-energy states are filled.  The
system is gapless if solution exists for $E=0$.  Since the
two terms inside the curly brackets in Eq. (\ref{spectrum}) are
positive definite, for $E=0$, both of them have to vanish, i.e.,
\begin{eqnarray}
J_z^2 &=& \Big[ J_x^2 + J_y^2 + 2 \cos k_1 J_x J_y \Big]^{\frac{1}{2}}
\nonumber \\ &&\times
\Big[ J_x^2 + J_y^2 + 2 \cos k_2 J_x J_y \Big]^{\frac{1}{2}}, 
\label{cond1}\\
\cos \phi_{\bf k} &=& 1. \label{cond2}
\end{eqnarray}
The values of $k_1$ and $k_2$ for which the gap vanishes are
determined by Eq.~(\ref{cond1}); $k_3$ is then given by
Eq.~(\ref{cond2}).  Solutions of Eq.~(\ref{cond1}) exist only when
\mbox{$J_z \le J_x + J_y$}, \mbox{$J_x \le J_y + J_z$} and \mbox{$J_y
  \le J_z + J_x$}; these conditions are same as that for the 2D Kitaev
model. Figure \ref{fermisurface} shows the plot of contours satisfying
Eq.~(\ref{cond1}) projected on to $k_3=0$ plane, where we have set
$J_x=J_y=1$ and varied $J_z$ from zero to two. The contour shrinks to the
point $(0,0)$ as $J_z$ approaches $J_x+J_y=2$, i.e., when the gap
opens up.

\begin{center}
\begin{figure}
\includegraphics[width=.35\textwidth]{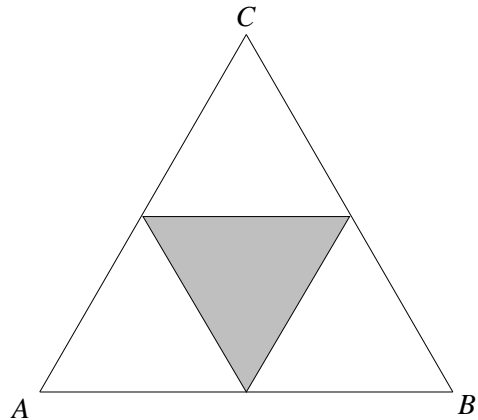}
\caption{ The phase diagram: it shows the plane defined by \mbox{$J_x
    +J_y+J_z =1$}. Point $A$ corresponds to \mbox{$J_x=1$} and
  \mbox{$J_y=J_z=0$}, $B$ corresponds to \mbox{$J_y=1$} and
  \mbox{$J_z=J_x=0$}, and $C$ corresponds to \mbox{$J_z=1$} and
  \mbox{$J_x=J_y=0$}. The shaded region is the gapless phase.}
\label{phasediagram}
\end{figure}
\end{center}

Figure \ref{phasediagram} depicts the phase diagram on a section of the
parameter phase. Interestingly, the phase diagram is symmetric in the
three coupling constants though they do not appear symmetrically in
the Hamiltonian (the $z$-links cannot be transformed to $x$ or $y$
links by any symmetry transformation of the lattice). Nor is the
spectrum, given by Eq. (\ref{spectrum}), symmetric in the coupling
constants. 

We end the discussion on the ground state by noting that spin
correlations can be calculated in exactly the same way as in
2D.\cite{BaskaranMandal,ChenNussinov} The only non-zero correlations
are those constructed out of the interaction terms that appear in the
Hamiltonian.

\section{\label{sec-excitations} Flux excitations}

The Kitaev model in 2D has anyonic excitations, i.e., there are
particle-like excitations which obey nontrivial statistics. Anyons are
very specific to 2D and cannot exist in higher dimensions; the
fundamental reason being that in $D > 2$ there are no nontrivial paths
which take a particle around another. However, in 3D, excitations
localized on loops may obey nontrivial statistics. In this section, we
show that there are excitations in our model which are localized on
loops. However, we do not address the issue of their statistics in
this paper.

We have seen that in the ground state $W_p = 1$ for all plaquettes
$p$. It then follows that the excitations are of two types:
(1) Flux configurations which violate the condition $W_p
  = 1$, i.e., $W_p = -1$ for some of the loops. It can be looked upon
  as creating a $\pi$ flux over those loops. (2) Fermionic excitations in the background of
  static configurations of $W_p$.

\begin{center}
\begin{figure}[h!]
\includegraphics[width=.4\textwidth]{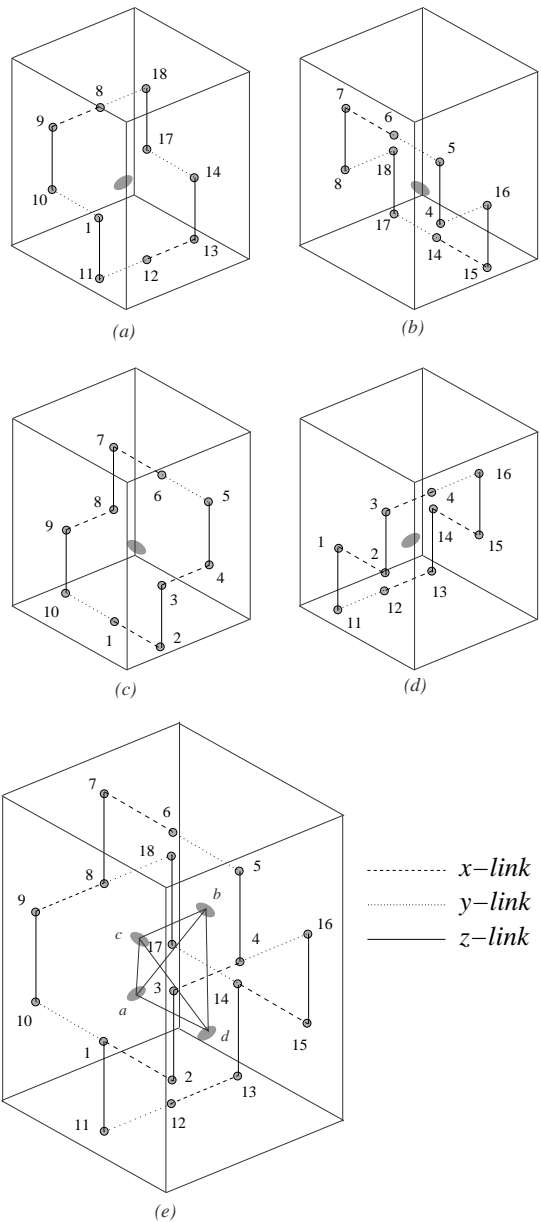}
\caption{[(a)-(d)] The four types of plaquettes. (e) Part of the lattice
  involving four such adjacent plaquettes; the corresponding operators give
  rise to a constraint. The ellipses, labeled $a$-$d$, respectively,
  represent each of the loops and form a ``tetrahedron''.}
\label{fig-loops}
\end{figure}
\end{center}

We will next show that the excitations of the first type have the
structure of loops.  Earlier we mentioned that not all $W_p$ are
independent; now we will find the constraints among them. There are
four types of plaquettes which are not related by translation. Let
$a,b,c,$ and $d$ be the labels for the different types [see Figs.
\ref{fig-loops}(a)-\ref{fig-loops}(d)].  Consider a part of the
lattice, shown in Fig.~\ref{fig-loops}(e), which consists of four
adjacent plaquettes---each a different type. Here the links are
labeled 1-20.  Let the corresponding loop operators be $W_a, W_b,
W_c,$ and $W_d$, respectively. In terms of the spin variables these
operators are
\begin{eqnarray}
  W_a &=&  \sigma^x_{11} \sigma^z_{12} \sigma^y_{13} \sigma^x_{14}
  \sigma^x_{17} \sigma^x_{18} \sigma^z_{8} \sigma^y_{9} \sigma^x_{10}
  \sigma^x_{1},  \nonumber \\   
  W_b &=&  \sigma^z_{14} \sigma^y_{15} \sigma^x_{16} \sigma^x_{4} \sigma^x_{5}
  \sigma^z_{6} \sigma^y_{7} \sigma^x_{8} \sigma^x_{18} \sigma^x_{17},
  \nonumber\\   
  W_c &=&  \sigma^z_{1} \sigma^y_{2} \sigma^y_{3} \sigma^y_{4} \sigma^x_{5}
  \sigma^z_{6} \sigma^y_{7} \sigma^y_{8} \sigma^y_{9} \sigma^x_{10}, 
  \nonumber\\ 
  W_d &=&  \sigma^x_{11} \sigma^z_{12} \sigma^y_{13} \sigma^y_{14}
  \sigma^y_{15} \sigma^x_{16} \sigma^z_{4} \sigma^y_{3} \sigma^y_{2} 
  \sigma^y_{1}.
\end{eqnarray}
Using the relations $\sigma_j^x \sigma_j^y \sigma_j^z = i$ and ${\sigma_j^a}^2=1$,
\begin{equation}
W_a W_b W_c W_d = 1.
\label{constraint}
\end{equation}

The above constraint has a graphical interpretation.  Note that a loop
operator can also be written as a product of the interaction terms on
the links contained in the loop.  In Fig.~\ref{fig-loops}, the object
(e), which is obtained by putting together the four loops (a)-(d),
represents the left-hand side of Eq.~(\ref{constraint}). Evidently,
each link in (e) is shared by two of the $W_p$'s. Since the square of
any interaction term is one, Eq.  (\ref{constraint}) immediately
follows.

\begin{center}
\begin{figure}
\includegraphics[width=.4\textwidth]{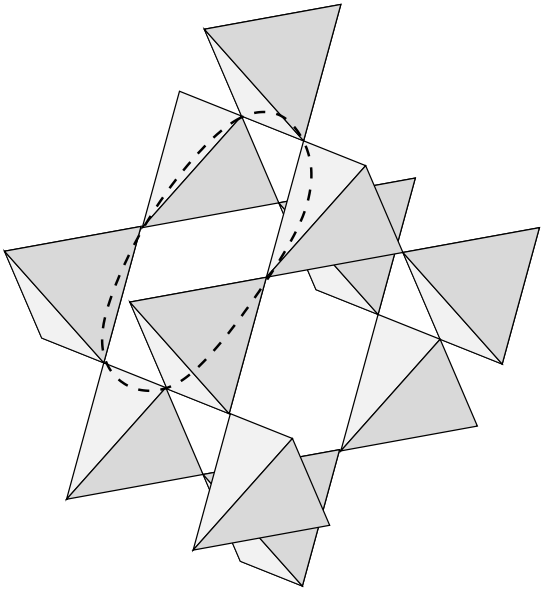}
\caption{ The lattice $\mathfrak{L}$ formed by the plaquettes---the
  pyrochlore lattice. Loops such as the dashed one, which goes through
  six sites, are the shortest (apart from the basic triangles).}
\label{fig-pyrochlore}
\end{figure}
\end{center}
For open boundary conditions, relations such as Eq.~(\ref{constraint})
exhaust all the constraints.  To find the configurations of $\{W_p\}$
which are consistent with the constraints, it is instructive to
consider a lattice obtained by representing each plaquette by a single
site. A plaquette has a step like structure, consisting of two
rectangles perpendicular to the $x$-$y$ plane connected by another
rectangle on the $x$-$y$ plane. Each loop can be uniquely represented
by a point at the center of the rectangle on the $x$-$y$ plane [in
Fig.~\ref{fig-loops}(e)such points are marked by ellipses]. Let
$\mathfrak{L}$ be the new lattice thus obtained; topologically, it is
the pyrochlore lattice, an arrangement of corner-sharing tetrahedra
(see Fig.~\ref{fig-pyrochlore}). (The edges of the geometrical object
formed by the centers of four adjacent loops are not of equal length,
and hence do not form an exact tetrahedron but a stretched one.
However, the connectivity of $\mathfrak{L}$ is same as that of the
pyrochlore lattice, and we will continue to refer to the basic objects
as tetrahedron.)

In this description, the four plaquettes that give rise to the
constraint in Eq.~(\ref{constraint}) are the four sites of a
tetrahedron, and each tetrahedron corresponds to an independent
constraint. Therefore, any configuration satisfying all the
constraints will have an even number of $W_p$ taking value of $-1$ in
each tetrahedron, where $p$ is now the site index in $\mathfrak{L}$.
Now it is clear how to obtain such configurations: draw a loop
$\mathcal{C}$ which does not cross itself and which lies entirely
within the tetrahedra, and let
\begin{eqnarray*}
W_p&=& \left\{
\begin{array}{rl}
-1,& \textrm{ if } p \in \mathcal{C}, \\
1, & \textrm{ otherwise.}
\end{array}
\right.
\end{eqnarray*}
Any closed self-avoiding loop contains an even number of sites (0, 2,
or 4) belonging to any particular tetrahedron; hence all the
constraints are satisfied. In other words, the flux excitations have
the structure of loops in the lattice $\mathfrak{L}$.

\section{\label{sec-heff}Effective Hamiltonian in the large $J_z$
  limit}
\begin{center}
\begin{figure}
\includegraphics[width=.4\textwidth]{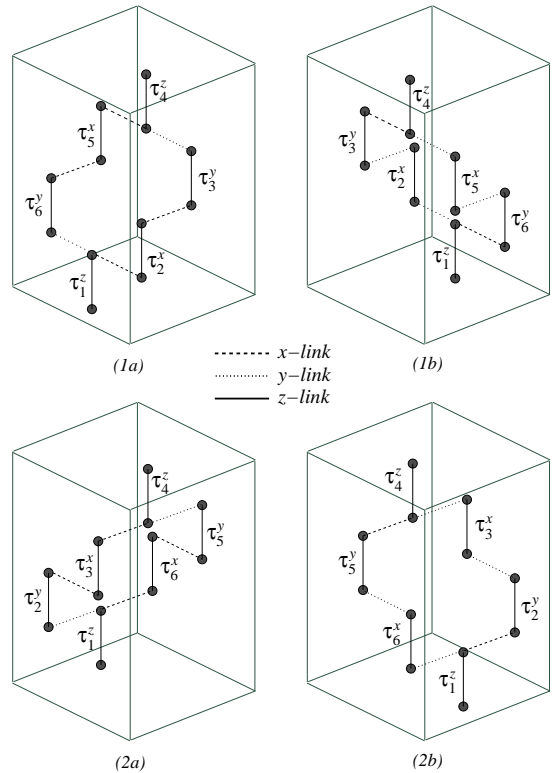}
\caption{\label{fig-heffloops} Representation of the flux operators in
  $H_{\rm{eff}}$ associated with the four types of plaquettes.}
\end{figure}
\end{center}

To study the excitations in the gapped phase, one can follow
Kitaev's\cite{Kitaev2} approach to the 2D model---a perturbative
analysis in the limit $J_z\gg J_x,J_y$. The unperturbed
Hamiltonian $H_0$ and the perturbation $H_1$ are
\begin{eqnarray}
H_0 &=& -J_z \sum_{<ij>_z} \sigma_i^z \sigma_j^z, \\
H_1 &=& -J_z \sum_{<ij>_x} \sigma_i^x \sigma_j^x 
-J_z \sum_{<ij>_y} \sigma_i^y \sigma_j^y, \\
H&=& H_0 + H_1 .
\label{unpertham} 
\end{eqnarray}
The ground state of $H_0$ is highly degenerate: the two spins on each
of the $z$-links can be either $|\uparrow \uparrow \rangle$ or
$|\downarrow \downarrow \rangle$. This degeneracy is lifted only at
the sixth order of perturbation theory. We now calculate the effective
low-energy Hamiltonian to this order.

The groundstate subspace of
$H_0$ can be thought of as the Hilbert space of effective
spin $\frac{1}{2}$'s located at each $z$ link. Let
$\tau_m^x, \tau_m^y$ and $\tau_m^z$ be Pauli matrices acting on the
effective spin at the $z$ link denoted by $m$, such that,
\begin{eqnarray}
\tau^z_m   |\uparrow \uparrow \rangle_m &=& |\uparrow \uparrow \rangle_m,
\nonumber \\
\tau^z_m |\downarrow \downarrow \rangle_m &=& -|\downarrow \downarrow
\rangle_m.
\end{eqnarray} 

A plaquette consists of four $z$-links and six $x$ or $y$-links. 
Consider a plaquette by also including the two $z$-links that
are directed out of the loop (see Fig. \ref{fig-heffloops}). Let us label the
$z$-links in such an object with an index $i = 1,2,\dots, 6$. The
labeling scheme for the four different types of loops is shown in the
figure. We associate with each plaquette $p$ a flux operator
$B_p$ defined as follows.
\begin{equation}
B_p = \tau_{p1}^z \tau_{p2}^{a_2} \tau_{p3}^{a_3} \tau_{p4}^z
\tau_{p5}^{a_5} \tau_{p6}^{a_6},
\end{equation}
where, for $i=2,3,5,6$, $a_i=x$ if the two links in the loop attached to
the $i$th $z$ link are the same; and $a_i = y$ otherwise. The flux
operator associated with the various types of loops are also shown in
Fig. \ref{fig-heffloops}.  
Then, to sixth order---which is the leading
order---the effective Hamiltonian, ignoring the constant terms arising at
the lower orders, is
\begin{equation}
H_{\rm{eff}} =  -\frac{7J_x^4J_y^2}{256J_z^5} \sum_{p}  B_{p}
-\frac{7J_y^4J_x^2}{256J_z^5} {\sum_{p}}^\prime  B_{p}, 
\label{heffect}     
\end{equation}
where the unprimed sum is over the loops of types $1a$ and $2a$ in Fig.
\ref{fig-heffloops}, which contain four $x$ links and two $y$ links,
while the primed sum is over loops of types $1b$ and $2b$, which
contain four $y$ links and two $x$ links.  Furthermore,
\begin{eqnarray}
[B_p, B_{p^\prime}] &=& 0, \label{bcomm} \\
B_p^2 &=& 1,~\forall p,p^\prime. \label{bpsquare} 
\end{eqnarray}
Thus, the operators $B_p$ can be simultaneously diagonalized for all
values of $p$ and their eigenstates will also be the
eigenstates of $H_{\rm{eff}}$. From Eq. (\ref{bpsquare}) it follows that
the eigenvalues of $B_p$ are $\pm 1$. Therefore, the ground state
$|GS\rangle_{\rm{eff}}$ will be such that 
\begin{equation}
B_p |GS\rangle_{\rm{eff}} = |GS\rangle_{\rm{eff}}, ~~ \forall p.
\label{bpgs}
\end{equation}

It is straightforward to see that $B_p$ is the projection of
$W_p$---the conserved quantities of the spin model defined in Eq.
(\ref{plaquetteop})---on the ground-state subspace of $H_0$. Therefore,
in the large $J_z$ limit, Eq. (\ref{bpgs}) is a confirmation of our
numerically verified assumption earlier that the ground state of $H$ is
vortex free.

The excitations are the states which violate Eq. (\ref{bpgs}). i.e.,
$B_p=-1$ for $p$ in some subset of plaquettes. However, not all
$B_p$ are independent---there exist constraints similar to Eq.
(\ref{constraint})---therefore the plaquettes with $B_p = -1$
cannot be chosen arbitrarily. To understand the properties of the
excitations, such as their statistics, a detailed analysis is
required. This is currently being done.

\section{\label{sec-summary}Summary and discussion}

We have constructed and solved a three-dimensional spin-$\frac{1}{2}$
model which is a generalization of the Kitaev model on a honeycomb
lattice. Based on the methods used by Kitaev,\cite{Kitaev2} we
calculated the exact low-energy fermionic spectrum by mapping the spin
model to one of free fermions in the background of a static
$\mathbb{Z}_2$ gauge field; the system has a gapped phase and a
gapless one. Quite interestingly, the gap vanishes on a contour in the
${\bf k}$ space; this could be related to some accidental
degeneracies---unrelated to any symmetry of the Hamiltonian---of the
ground state in the classical limit. The two-dimensional Kitaev model
has been shown to have such a degeneracy in the classical limit, which
grows exponentially with the system size.\cite{BaskaranSen} This
result can be readily generalized to our model. It will be interesting
to see how quantum fluctuations lifts this degeneracy.

We have further shown that the excitations of the gauge field, due to
some local constraints, have the topology of loops. As a first step
towards understanding the nature of the excitations in the gapped
phase, we have derived an effective $\mathbb{Z}_2$ gauge theory in the
limit $J_z\gg J_x,J_y$. The ground state of the effective Hamiltonian
thus obtained is trivially solved and it is consistent with the
assumption we made in the calculation of the fermionic spectrum that
the ground state is vortex free. Further study of the effective theory
should tell us more about the properties of the excitations, such as
whether they obey fractional statistics.

Yao and Kivelson\cite{YaoKivelson} have introduced a two-dimensional Kitaev model in
which one type of plaquettes is a triangle;
the existence of loops with odd number of links results in the
spontaneous breaking of time-reversal symmetry. We mention that in our
model also the time reversal symmetry can be similarly broken by
replacing each vertex in the lattice with a triangle.

 {\it{Note added}}~: Toward the completion of this paper, we came
 across the work by Si and Yu in Ref. \onlinecite{YuSi} which
 discusses a variety of exactly solvable Kitaev models in three
 dimensions, none of which are identical to the model introduced here.

\acknowledgments

We thank G. Baskaran and R. Shankar for very useful discussions.

\end{document}